\documentclass[12pt]{article}

\newcommand{\pslash}{\not \! p}

\newcommand{\delslash}{\not \! \partial}

\usepackage{amsmath,amssymb,amscd,amsbsy,amsgen,amsopn,amstext,
amsxtra}

\usepackage[dvips]{graphicx}

\begin{document}

\begin{flushright}

\end{flushright}

\vskip 0.5 truecm

\begin{center}
{\Large{\bf Electromagnetic interaction in theory with Lorentz invariant CPT violation}}
\end{center}
\vskip .5 truecm
\begin{center}
{\bf { Masud Chaichian{$^*$}, Kazuo Fujikawa$^\dagger$ and
Anca Tureanu$^*$}}
\end{center}

\begin{center}
\vspace*{0.4cm} {\it { $^*$Department of Physics, University of
Helsinki, P.O.Box 64, FIN-00014 Helsinki,
Finland\\
$^\dagger$ Mathematical Physics Laboratory, RIKEN Nishina Center,\\
Wako 351-0198, Japan}}
\end{center}


\begin{abstract}
An attempt is made to incorporate the electromagnetic interaction in a
  Lorentz invariant but CPT violating non-local model with
particle-antiparticle mass splitting, which is regarded as a modified
QED. The gauge invariance is maintained by the Schwinger
non-integrable phase factor but the electromagnetic interaction breaks
C, CP and CPT symmetries. Implications of the present CPT breaking
scheme on the
electromagnetic transitions and particle-antiparticle pair creation
are discussed. The CPT violation such as the one suggested here may open a new path
 to the analysis of baryon asymmetry  since some of the Sakharov constraints
 are expected to be modified.
\end{abstract}


\section{Introduction}
The local field theory defined in Minkowski space-time is very
successful, and CPT symmetry is a fundamental symmetry of any such
theory~\cite{pauli}. Nevertheless, the possible breaking of CPT
symmetry has also been discussed. One of the logical ways to break CPT
symmetry is to make the theory non-local by preserving Lorentz
symmetry, while the other is to break Lorentz symmetry itself. The
Lorentz symmetry breaking scheme
has been mainly studied in the past, including its physical
implications~\cite{ellis, piguet}. A proposal of Lorentz invariant CPT
breaking scheme is relatively new~\cite{chaichian} and its logical
consistency has also been emphasized~\cite{JGB}. But its physical
implications have not been analyzed except for the recent proposal of
an explicit Lagrangian model of particle--antiparticle mass
splitting~\cite{chaichian2} and its application to the neutrino--antineutrino mass splitting in the Standard Model~\cite{chaichian3}.
It was emphasized there that only the neutrino mass terms in the
Standard Model can preserve the basic local
$SU(2)_{L}\times U(1)$ gauge symmetry in the Lorentz invariant
non-local CPT breaking scheme without introducing non-integrable phase
factors. From the point of view of particle phenomenology, this
uniqueness of the neutrino mass splitting in the Standard Model is
quite interesting~\cite{murayama, altarelli, adamson}.

If one wants to accommodate the non-local Lorentz invariant CPT
breaking mechanism
in the couplings of  general elementary particles, one needs to go
beyond the conventional local gauge principle by incorporating the
Schwinger non-integrable phase factor. (This non-integrable phase
factor is also known as the Wilson-line integral in lattice gauge
theory, and we use the terms Schwinger's factor, non-integrable
phase factor and Wilson-line interchangeably in the present Letter.)

To be specific, we adopt the simplest Lorentz invariant and non-local
CPT breaking Hermitian Lagrangian~\cite{chaichian2}:
\begin{eqnarray}\label{(2.3)}
S&=&\int d^{4}x\Big\{\bar{\psi}(x)i\gamma^{\mu}\partial_{\mu}\psi(x)
  - m\bar{\psi}(x)\psi(x)\\
  && -i\mu\int
d^{4}y[\theta(x^{0}-y^{0})-\theta(y^{0}-x^{0})]\delta((x-y)^{2}-l^{2})[\bar{\psi}(x)\psi(y)]\Big\},\nonumber
\end{eqnarray}
as a model Lagrangian of a charged fermion ("electron") and study its
electromagnetic interactions.
For the real parameter $\mu$,
the third term has C = CP = CPT = $-1$ and thus no symmetry to ensure the equality
of particle and antiparticle masses. The dimension of $\mu$ depends on
the choice of the non-local factor $\delta((x-y)^{2}-l^{2})$ and in
the present case
it is $[M]^{3}$, while $l$ has dimension of length.

The free equation of motion for the fermion is 
\begin{eqnarray}\label{(2.4)}
&&i\gamma^{\mu}\partial_{\mu}\psi(x)=m\psi(x)\\
&&+i\mu\int
d^{4}y[\theta(x^{0}-y^{0})-\theta(y^{0}-x^{0})]\delta((x-y)^{2}-l^{2})\psi(y).\nonumber
\end{eqnarray}
By inserting an Ansatz for the possible solution,
$\psi(x)=e^{-ipx}U(p)$,
we obtain
\begin{eqnarray}\label{(2.6)}
\pslash U(p)&=&mU(p)
+i\mu[f_{+}(p)-f_{-}(p)]U(p),
\end{eqnarray}
where $f_{\pm}(p)$ is a Lorentz invariant "form factor" defined by
\begin{eqnarray}\label{(1.3)}
&&f_{\pm}(p)=\int d^{4}z_{1}
e^{\pm ipz_{1}}\theta(z_{1}^{0})\delta((z_{1})^{2}-l^{2}),
\end{eqnarray}
which are  inequivalent for time-like $p$ due to the factor
$\theta(z_{1}^{0})$.
By assuming a time-like $p$, we go to the frame where $\vec{p}=0$.
Then the eigenvalue equation for the mass is given by~\footnote{
  It is possible to assign a finite value to the last term in eq. (5)
for $p_{0}\neq 0$ by using the formal relation,
\begin{eqnarray}
\int_{0}^{\infty}dz\frac{z^{2}\sin [
p_{0}\sqrt{z^{2}+l^{2}}]}{\sqrt{z^{2}+l^{2}}}&=&-\frac{\partial^{2}}{\partial
p_{0}^{2}}\int_{0}^{\infty}dz\frac{z^{2}\sin
[p_{0}\sqrt{z^{2}+l^{2}}]}{[z^{2}+l^{2}]^{3/2}}.\nonumber
\end{eqnarray}
}

\begin{eqnarray}\label{(2.10)}
p_{0}&=&\gamma_{0}\left[m - 4\pi \mu\int_{0}^{\infty}dz\frac{z^{2}\sin [
p_{0}\sqrt{z^{2}+l^{2}}]}{\sqrt{z^{2}+l^{2}}}\right],
\end{eqnarray}
where we used  the explicit formula
\begin{eqnarray}\label{(1.4)}
f_{\pm}(p^{0})
&=&2\pi \int_{0}^{\infty}dz\frac{z^{2}e^{\pm
ip^{0}\sqrt{z^{2}+l^{2}}}}{\sqrt{z^{2}+l^{2}}}.
\end{eqnarray}
This eigenvalue equation under $p_{0}\rightarrow -p_{0}$ becomes
(after sandwiching by $\gamma_{5}$)
\begin{eqnarray}\label{(2.13)}
p_{0}
&=&\gamma_{0}\left[m + 4\pi\mu\int_{0}^{\infty}dz\frac{z^{2}\sin [
p_{0}\sqrt{z^{2}+l^{2}}]}{\sqrt{z^{2}+l^{2}}}\right],
\end{eqnarray}
which is not identical to the original equation in \eqref{(2.10)}.
This causes the mass splitting of particle and antiparticle in the
sense of Dirac.
One may
solve the mass eigenvalue equations iteratively by assuming that the
terms with
the parameter $\mu$ are much smaller than $m$. One then obtains
the mass eigenvalues at
\begin{eqnarray}\label{(8)}
m_{\pm}
&\simeq&m \pm
4\pi\mu\int_{0}^{\infty}dz\frac{z^{2}\sin [
m\sqrt{z^{2}+l^{2}}]}{\sqrt{z^{2}+l^{2}}},
\end{eqnarray}
where the upper two (positive) components of the matrix
$\gamma_{0}$ in (5) and (7) are used.
See Ref.~\cite{chaichian2} for further details.

\section{Electromagnetic interaction -- modified QED}

To introduce the electromagnetic interaction in (1), we consider the
simplest scheme (a modified QED):
\begin{eqnarray}\label{(2.3)}
S&=&\int d^{4}x\Big\{\bar{\psi}(x)i\gamma^{\mu}D_{\mu}\psi(x)
  - m\bar{\psi}(x)\psi(x)\nonumber\\
  && -i\mu\int
d^{4}y[\theta(x^{0}-y^{0})-\theta(y^{0}-x^{0})]\delta((x-y)^{2}-l^{2})\nonumber\\
&&\times \bar{\psi}(x)\exp\left [ie\int_{y}^{x}A_{\mu}(z)dz^{\mu}\right]\psi(y)\Big\}
  \nonumber\\
  &&-\frac{1}{4}\int d^{4}xF_{\mu\nu}(x)F^{\mu\nu}(x),
\end{eqnarray}
with
\begin{eqnarray}
D_{\mu}=\partial_{\mu}-ieA_{\mu}(x).
\end{eqnarray}
We added the Schwinger non-integrable phase factor
\begin{eqnarray}
\exp\left[ie\int_{y}^{x}A_{\mu}(z)dz^{\mu}\right],
\end{eqnarray}
to make the non-local term gauge invariant.
This action is invariant under the gauge transformation
\begin{eqnarray}
&&\psi(x)\rightarrow e^{i\alpha(x)}\psi(x),\nonumber\\
&&A_{\mu}(x)\rightarrow A_{\mu}(x)+\frac{1}{e}\partial_{\mu}\alpha(x),
\end{eqnarray}
and the C, CP and CPT transformation properties of each term in the
action (8) are the same as in the theory without electromagnetic couplings.

It is natural to consider the non-integrable phase factor
in (8) as an independent dynamical entity rather than a given external
factor. In fact, Y. Nambu emphasized in many occasions the non-integrable
phase factor as
a manifestation of string-like objects appearing in the theory.

Our proposal here is to replace the non-integrable phase factor in (8)
by a first quantized very massive particle propagation defined by the
covariant path integral
\begin{eqnarray}
&&\exp\left [ie\int_{y}^{x}A_{\mu}(z)dz^{\mu}\right]\delta_{\alpha,\beta}\Rightarrow\\
&&\int{\cal
D}z^{\mu}\exp\Big\{i\int_{y}^{x}\frac{1}{2}\left[(\dot{z}^{\mu})^{2}+M^{2}\right]d\tau+ie\int_{y}^{x}A_{\mu}(z)\frac{dz^{\mu}}{d\tau}d\tau\Big\}\delta_{\alpha,\beta},
\nonumber
\end{eqnarray}
where the factor $\delta_{\alpha,\beta}$ contracts the spinor indices,
an analogue of the Chan-Paton factor in string theory.
In this way, the non-integrable phase factor becomes more dynamical
and the flow of the charge is visualized in Feynman diagrams, although
the second quantized particle and the first quantized particle appear
in a mixed manner in Feynman diagrams. This use of a semi-static
massive particle for the non-integrable phase
factor is common in lattice gauge theory.

As for the quantization of the theory non-local in time, we adopt the path
integral on the basis of Schwinger's action principle, which is
based on the equations of motion~\cite{fujikawa}.

\section{Current conservation and Ward--Takahashi identity}
One may examine the fermion pair creation through the lowest order
electromagnetic interaction, for example,  to study the implications
of the fermion and antifermion mass splitting on the pair production.
To the lowest order in $O(e)$, one may expand the non-integrable phase
factor as
$$\exp\left[ie\int_{y}^{x}A_{\mu}(z)dz^{\mu}\right]=1+ie\int_{y}^{x}A_{\mu}(z)dz^{\mu}.$$
The interaction part for the lowest order pair creation is given by
\begin{eqnarray}
S_{I}&=&e \int d^{4}x \bar{\psi}(x)\gamma^{\mu}A_{\mu}(x)\psi(x)\nonumber\\
&+&e\mu\int  d^{4}x
d^{4}y[\theta(x^{0}-y^{0})-\theta(y^{0}-x^{0})]\delta((x-y)^{2}-l^{2})\nonumber\\
&&\times\bar{\psi}(x)\left[\int_{y}^{x}A_{\mu}(z)dz^{\mu}\right]\psi(y)
\end{eqnarray}
and the electromagnetic current is
\begin{eqnarray}
J^{\mu}(w)&=&\frac{\delta}{\delta A_{\mu}(w)}S_{I}\nonumber\\
&=&e \bar{\psi}(w)\gamma^{\mu}\psi(w)\nonumber\\
&+&e\mu\int  d^{4}x
d^{4}y[\theta(x^{0}-y^{0})-\theta(y^{0}-x^{0})]\delta((x-y)^{2}-l^{2})\nonumber\\
&&\times\bar{\psi}(x)\int_{\tau_{y}}^{\tau_{x}}\left[\delta^{4}(z(\tau)-w)\frac{dz^{\mu}}{d\tau}\right]d\tau\,\psi(y),
\end{eqnarray}
where $z^{\mu}(\tau)$ stands for the coordinate of the massive particle.
The current conservation condition becomes
\begin{eqnarray}
\partial_{\mu}J^{\mu}(w)
&=&e \partial_{\mu}[\bar{\psi}(w)\gamma^{\mu}\psi(w)]\nonumber\\
&+&e\mu\int  d^{4}x
d^{4}y[\theta(x^{0}-y^{0})-\theta(y^{0}-x^{0})]\delta((x-y)^{2}-l^{2})\nonumber\\
&&\times\bar{\psi}(x)\int_{\tau_{y}}^{\tau_{x}}\left[\frac{\partial}{\partial
w^{\mu}}\delta^{4}(z(\tau)-w)\frac{dz^{\mu}}{d\tau}\right]d\tau\,\psi(y)\\
&=&e[\bar{\psi}(w)\delslash\psi(w)]-e(-\partial_{\mu}\bar{\psi}(w)\gamma^{\mu})\psi(w)\nonumber\\
&-&e\mu\int  d^{4}x
d^{4}y[\theta(x^{0}-y^{0})-\theta(y^{0}-x^{0})]\delta((x-y)^{2}-l^{2})\nonumber\\
&&\times\bar{\psi}(x)[\delta^{4}(x-w)-\delta^{4}(y-w)]\psi(y)\nonumber\\
&=&e[\bar{\psi}(w)\delslash\psi(w)]-e(-\partial_{\mu}\bar{\psi}(w)\gamma^{\mu})\psi(w)\nonumber\\
&-&e\mu\int
d^{4}y[\theta(w^{0}-y^{0})-\theta(y^{0}-w^{0})]\delta((w-y)^{2}-l^{2})\bar{\psi}(w)\psi(y)\nonumber\\
&+&e\mu\int  d^{4}x
[\theta(x^{0}-w^{0})-\theta(w^{0}-x^{0})]\delta((x-w)^{2}-l^{2})\bar{\psi}(x)\psi(w),\nonumber
\end{eqnarray}
which in fact vanishes if one uses the free equation of motion for the
fermion in (2) and its conjugate. Here we used the relation
\begin{eqnarray}
\frac{\partial}{\partial w^{\mu}}\delta^{4}(z(\tau)-w)\frac{dz^{\mu}}{d\tau}=
-\frac{d}{d\tau}\delta^{4}(z(\tau)-w).
\end{eqnarray}

If one remembers the inclusion of the path integral for the (free)
massive particle, one should actually write in the current (15)
\begin{eqnarray}
\delta^{4}(z(\tau)-w)\frac{dz^{\mu}}{d\tau}&\Rightarrow&\frac{1}{Z}\int
d^{4}z^{\prime}\delta^{4}(z^{\prime}(\tau)-w)\langle
x,\tau_{x}|z^{\prime}, \tau\rangle
\frac{d{z^{\prime}}^{\mu}}{d\tau}\langle z^{\prime},
\tau|y,\tau_{y}\rangle\nonumber\\
&=&\frac{1}{Z}\langle
x,\tau_{x}|\delta^{4}(\hat{z}(\tau)-w)\frac{d}{d\tau}\hat{z}^{\mu}(\tau)|y,\tau_{y}\rangle,
\end{eqnarray}
where $Z=\langle x,\tau_{x}|y,\tau_{y}\rangle$ is the normalization
factor of the path integral partition function, and the last
expression is given in the operator notation in the interaction
picture, since we are working in the lowest order of the electromagnetic coupling in (14) in the first quantized path integral. Note that
\begin{eqnarray}
&&\hat{z}^{\mu}(\tau)=e^{i\hat{H}_{0}\tau}\hat{z}^{\mu}(0)e^{-i\hat{H}_{0}\tau},
\nonumber\\
&&\hat{H}_{0}=\frac{1}{2}[\hat{P}^{2}_{\mu}(0)-M^{2}],
\end{eqnarray}
with $[\hat{P}_{\mu}(0), \hat{z}^{\nu}(0)]=i\hbar g_{\mu}^{\ \nu}$.
With the replacement in (18), one still obtains the same conservation
relation as in (16) if one notes the relation
\begin{eqnarray}
\delta^{4}(\hat{z}(\tau_{y})-w)|y,\tau_{y}\rangle=\delta^{4}(y-w)|y,\tau_{y}\rangle,
\end{eqnarray}
for example.

The first term in (14) is invariant under C (and in fact under CP
and CPT), while the second term is confirmed to be odd under the
conventional charge conjugation symmetry C (and in fact odd under
CP and CPT also). Thus the electromagnetic interaction breaks
those basic symmetries slightly. Nevertheless, this interaction is
invariant under the gauge transformation
$A_{\mu}(x)\rightarrow A_{\mu}(x)+\frac{1}{e}\partial_{\mu}\alpha(x)$
if one uses the equation of motion for the free fermion field in (2)
and its conjugate, as was already explained. This gauge invariance
ensures the Ward--Takahashi identity for the three-point vertex function
in the form
\begin{eqnarray}
\langle
T^{\star}\psi(u)S_{I}(A_{\mu}=\frac{1}{e}\partial_{\mu}\alpha)\bar{\psi}(w)\rangle
&=\int d^{4}x\,\langle
T^{\star}\psi(u)\bar{\psi}(x)\rangle\frac{1}{e}\alpha(x)\delta(x-w)\nonumber\\
&-\int d^{4}x\,\delta(u-x)\frac{1}{e}\alpha(x)\langle
T^{\star}\psi(x)\bar{\psi}(w)\rangle
\end{eqnarray}
in the interaction picture, since the interaction part $S_{I}$ in (14) is defined in the lowest order in the electromagnetic coupling and the current conservation condition (16) is satisfied by using the free equations of motion of the fermion field. The free propagator is defined by the
inverse of the
free equation of motion in (2), namely,
\begin{eqnarray}
&&\langle
T^{\star}\psi(x)\bar{\psi}(y)\rangle=\int\frac{d^{4}p}{(2\pi)^{4}}
e^{-ip(x-y)}\frac{i}{\pslash-m +i\epsilon-i\mu[f_{+}(p)-f_{-}(p)]}.
\end{eqnarray}

The full set of Ward-Takahashi identities, i.e. the relations among different  Green's functions for the presented modified QED, can be derived 
formally in the present path integral quantization, but the exact current becomes more involved than (15) due to the presence of the non-integrable phase factor in the full action (9). An analysis of higher order effects in the electromagnetic coupling in the presence of the non-integrable phase factor even in the lowest order of the CPT-violation parameter $\mu$ is an interesting subject of future study.

\section{Fermion pair creation}
We now consider the charged particle pair creation from a virtual photon
\begin{eqnarray}
\gamma (k) \rightarrow e (p)+\bar{e} (\bar{p}).
\end{eqnarray}
The current matrix element in the momentum space is given by using the
current in (15)
\begin{eqnarray}
J^{\mu}(k)&=&\int d^{4}w e^{-ikw}\langle
p,\bar{p}|J^{\mu}(w)|0\rangle\nonumber\\
&=&(2\pi)^{4}\delta(k-p-\bar{p})e\bar{u}(p)\gamma^{\mu}v(\bar{p})\nonumber\\
&+&e\mu\int  d^{4}x
d^{4}ye^{i\bar{p}y+ipx}[\theta(x^{0}-y^{0})-\theta(y^{0}-x^{0})]\delta((x-y)^{2}-l^{2})\nonumber\\
&&\times\bar{u}(p)\int_{\tau_{y}}^{\tau_{x}}\left[e^{-ikz(\tau)}\frac{dz^{\mu}}{d\tau}\right]d\tau\,
v(\bar{p}),
\end{eqnarray}
where we used the solutions $u(p)$ and $v(\bar{p})$ of the modified
Dirac equation (3) with masses in (8), and  the representation
\begin{eqnarray}
\delta^{4}\left(z(\tau)-w\right)=\int
\frac{d^{4}q}{(2\pi)^{4}}e^{iq\left[z(\tau)-w\right]}.
\end{eqnarray}
Here we use the original expression of the current in (15) without the
quantum fluctuation of the non-integrable phase factor; this is
because the dependence of the path integral normalization factor
$Z=\langle x,\tau_{x}|y,\tau_{y}\rangle$ on
the coordinates of the end-points complicates the evaluation, although it does
not make it impossible.
We thus employ the straight-line path between the two end-points:
\begin{eqnarray}
z^{\mu}(\tau)=(x^{\mu}-y^{\mu})\frac{\tau-
\tau_{y}}{\tau_{x}-\tau_{y}}+y^{\mu}
\end{eqnarray}
and we evaluate
\begin{eqnarray}
&&\int  d^{4}x
d^{4}ye^{i\bar{p}y+ipx}[\theta(x^{0}-y^{0})-\theta(y^{0}-x^{0})]\delta((x-y)^{2}-l^{2})\\
&&\times\int_{\tau_{y}}^{\tau_{x}}d\tau\exp\{-i[(kx-ky)\frac{\tau-
\tau_{y}}{\tau_{x}-\tau_{y}}+ky]\}\frac{(x^{\mu}-y^{\mu})}{\tau_{x}-\tau_{y}}\nonumber\\
&=&\int_{0}^{1}d\eta\int  d^{4}x
d^{4}ye^{i\bar{p}y+ipx}[\theta(x^{0}-y^{0})-\theta(y^{0}-x^{0})]\delta((x-y)^{2}-l^{2})\nonumber\\
&&\times\exp\{-i[k(x-y)\eta+ky]\}(x^{\mu}-y^{\mu})\nonumber\\
&=&-i\left(\frac{\partial}{\partial p_{\mu}}-\frac{\partial}{\partial
\bar{p}_{\mu}}\right)\int_{0}^{1}d\eta\int  d^{4}x
d^{4}ye^{i\bar{p}y+ipx}[\theta(x^{0}-y^{0})-\theta(y^{0}-x^{0})]\nonumber\\
&&\times\delta((x-y)^{2}-l^{2})\exp\{-i[k(x-y)\eta+ky]\}\nonumber\\
&=&-i(2\pi)^{4}\delta(k-p-\bar{p})\left(\frac{\partial}{\partial
p_{\mu}}-\frac{\partial}{\partial
\bar{p}_{\mu}}\right)\int_{0}^{1}d\eta\nonumber\\
&&\times\int  d^{4}u
[\theta(u^{0})-\theta(-u^{0})]\delta(u^{2}-l^{2})
\exp\{-iku(\eta-1)-i\bar{p}u\}\nonumber\\
&=&(2\pi)^{4}\delta(k-p-\bar{p})\left(-i\frac{\partial}{\partial
\bar{p}_{\mu}}\right)\int_{0}^{1}d\eta[f_{+}\left(k(\eta-1)+\bar{p}\right)-f_{-}\left(k(\eta-1)+\bar{p}\right)],\nonumber
\end{eqnarray}
where we defined $\eta=\frac{\tau- \tau_{y}}{\tau_{x}-\tau_{y}}$ and
$u=x-y$, and used the form factor defined in (4).
We thus have the current (by suppressing the factor
$(2\pi)^{4}\delta(k-p-\bar{p})$)
\begin{eqnarray}
J^{\mu}(k)&=&e\bar{u}(p)\gamma^{\mu}v(\bar{p})+e\mu\bar{u}(p)F^{\mu}(p,
\bar{p})v(\bar{p})
\end{eqnarray}
with
\begin{eqnarray}
F^{\mu}(p, \bar{p})\equiv\Big\{(-i\frac{\partial}{\partial
\bar{p}_{\mu}})\int_{0}^{1}d\eta\left[f_{+}\left(k(\eta-1)+\bar{p}\right)-f_{-}\left(k(\eta-1)+\bar{p}\right)\right]\Big\}|_{k=p+\bar{p}}.
\end{eqnarray}
We have a small correction $F^{\mu}(p, \bar{p})$ to the electromagnetic
current,  which flips chirality (and thus it is similar to the Pauli term)
and violates C, CP and CPT. Note that the first term in (28)
alone is not conserved due to the mass splitting, but the first and
second terms in (28) put together are conserved, as eq. (16) indicates.

\section{Discussion}
It is interesting that the gauge invariance is maintained in Lorentz
invariant non-local theory (9)
by a scheme apparently different from that in local theory.
The crucial point is that the equality of  masses does not play any
essential role in this analysis of gauge invariance. This gauge
invariance is somewhat analogous to the gauge invariance of the Pauli
term in the ordinary coupling of the photon to charged fermions.
Another interesting aspect is that the fermion line is not continuous
in space-time in the non-local term, as is seen from the expression of  $S_{I}$ in (14),
which is something new in view of the Feynman's picture of charged
particle propagation in space-time. To reconcile this discontinuous
world-line in space-time with Feynman's picture, we have suggested
that the non-integrable phase factor represents a {\em very massive
semi-static fermion} propagation in the first quantization picture,
which is common in lattice gauge theory. In this way, one can maintain
the manifestly continuous flow of the charged particle and thus the
continuous flow of the electric current.

Physically one may argue that the presence of this (indefinite) very
massive particle in the intermediate stage of the flow of the charge
allows for the possible appearance of the mass difference between the
particle and antiparticle, which is consistent with gauge invariance,
in  our Lorentz invariant CPT breaking model.

As for practical implications of CPT breaking in the present modified
QED, the search for the mass splitting of particle and antiparticle,
just as the search for the neutrino antineutrino mass splitting in
oscillation experiments~\cite{adamson}, is interesting~\cite{Aaltonen, hori}.
In the atomic transitions of the matter or antimatter systems,  the
frequency differences caused by the small mass difference between the
"electron" and "positron" such as in (8) will be important.
Other possibilities are to look for the possible small C and CP breaking in electromagnetic interactions other than those caused by weak
interactions.  The effects of unitarity breaking are expected to be
minimal if one considers the processes lowest order in the small CPT
breaking non-local terms\footnote{To our knowledge, there exists no
example of a viable field theoretical model of elementary particles,
which breaks CPT invariance (in addition to C and CP breaking) while
preserving unitarity, regardless of whether Lorentz invariance is
violated or not. The analysis performed in~\cite{piguet}, which is
based on a local but both CPT and Lorentz invariance violating
modified  QED, indicates that unitarity is also generally violated when the
charge conjugation symmetry C is broken, the latter being an
experimental fact observed in weak interactions. Our present CPT
violating QED, which is Lorentz invariant but non-local, is expected
to break unitarity in general. Therefore, it may be natural to consider
the CPT violating theories, in particular the modified QED studied in
the present work, as effective theories emerging from a more
fundamental theory, for example in higher dimensions, and the unitarity 
issue may be disregarded in effective theories.}.

The analysis in this Letter suggests the following questions: 

{\it i)} The parameter $\mu$  in eqs. (1) and (9)
controls C, CP and CPT breaking, while Lorentz and gauge  invariance are
maintained. However,  C violation has not been observed in
electromagnetic interactions. What bound does this impose on the mass
splitting?

{\it ii)} Is this bound also reflected as a lower mass for the static heavy
fermion?

{\it iii)} Can this mechanism be used to generate the baryon asymmetry of the
Universe in equilibrium (with CPT violation, the three Sakharov
conditions can be
ignored ) and  what does such a case imply for the parameters of the model?

We thank the anonymous Referee for raising the above basic questions,
and we plan to address those issues in a future communication.

\section*{Acknowledgements}

The support of the Academy
of Finland under the Projects No. 136539 and 140886 is gratefully
  acknowledged.

\end{document}